\documentclass[final,5p,times,twocolumn]{elsarticle}

\usepackage{graphicx}
\usepackage{amssymb}
\usepackage{color}
\newcommand{\nc}{\newcommand}           
\nc{\vc}[1]     {\mbox{\boldmath $#1$}} 
\nc{\wtil}      {\widetilde}            
\nc{\bra}       {\langle}               
\nc{\ket}       {\rangle}               
\nc{\bras}[1]   {\langle#1|}            
\nc{\kets}[1]   {|#1\rangle}            
\nc{\hO}        {O}           

\def\JL#1#2#3#4{ {{\rm #1}} #2 (#3) #4}  
\nc{\PR}[3]     {\JL{Phys. Rev.}{#1}{#2}{#3}}
\nc{\PRC}[3]    {\JL{Phys. Rev.~C}{#1}{#2}{#3}}
\nc{\PRA}[3]    {\JL{Phys. Rev.~A}{#1}{#2}{#3}}
\nc{\PRL}[3]    {\JL{Phys. Rev. Lett.}{#1}{#2}{#3}}
\nc{\NP}[3]     {\JL{Nucl. Phys.}{#1}{#2}{#3}}
\nc{\NPA}[3]    {\JL{Nucl. Phys.}{A#1}{#2}{#3}}
\nc{\PL}[3]     {\JL{Phys. Lett.}{#1}{#2}{#3}}
\nc{\PLB}[3]    {\JL{Phys. Lett.~B}{#1}{#2}{#3}}
\nc{\PTP}[3]    {\JL{Prog. Theor. Phys.}{#1}{#2}{#3}}
\nc{\PTPS}[3]   {\JL{Prog. Theor. Phys. Suppl.}{#1}{#2}{#3}}
\nc{\PRep}[3]   {\JL{Phys. Rep.}{#1}{#2}{#3}}
\nc{\JP}[3]     {\JL{J. of Phys.}{#1}{#2}{#3}}
\nc{\andvol}[3] {{\it ibid.}\JL{}{#1}{#2}{#3}}
\nc{\mydraft}	{\setlength{\topmargin}{-1.5cm}}

\journal{Physics Letters B}
\begin{document}
\begin{frontmatter}

\title{Five-body resonances of $^8$He using the complex scaling method}

\author[a,b]{Takayuki Myo \corref{cor1}}
\cortext[cor1]{Corresponding author}
\ead{myo@ge.oit.ac.jp}
\address[a]{General Education, Faculty of Engineering, Osaka Institute of Technology, Osaka, Osaka 535-8585, Japan}
\address[b]{Research Center for Nuclear Physics (RCNP), Osaka University, Ibaraki 567-0047, Japan}
\author[c]{Ryosuke Ando}
\ead{ando@nucl.sci.hokudai.ac.jp}
\author[c]{Kiyoshi Kat\=o}
\ead{kato@nucl.sci.hokudai.ac.jp}
\address[c]{Division of Physics, Graduate School of Science, Hokkaido University, Sapporo 060-0810, Japan}

\begin{abstract}
The $0^+$ states of $^8$He are studied in a five-body $^4$He+$n$+$n$+$n$+$n$ cluster model.
Many-body resonances are treated on the correct boundary condition as Gamow states using the complex scaling method.
The $0^+_2$ state of $^8$He is predicted as a five-body resonance in the excitation energy of 6.3 MeV with a width of 3.2 MeV, 
which mainly has a $(p_{3/2})^2(p_{1/2})^2$ configuration.
In this state, number of the $0^+$ neuron pair shows almost two, which is different from the ground state having a large amount of the $2^+$ pair component.
The monopole transition of $^8$He from the ground state into the five-body unbound states is also evaluated. 
It is found that the $^7$He+$n$ component mostly exhausts the strength, while the $0^+_2$ contribution is negligible. 
The final states are dominated by $^6$He+$n$+$n$, not $^4$He+$n$+$n$+$n$+$n$.
The results indicate the sequential breakup process of $^8$He $\to$ $^7$He+$n$ $\to$ $^6$He+$n$+$n$ by the monopole excitation.
\end{abstract}


\end{frontmatter}


The development of experiments using radioactive beam has provided us 
with much information on unstable nuclei far from the stability.
In particular, the light nuclei near the drip-line exhibit new phenomena of nuclear structures,
such as the neutron halo structure in $^6$He, $^{11}$Li and $^{11}$Be \cite{tanihata85}.

Recently, many experiments on $^8$He have been reported \cite{iwata00,meister02,chulkov05,skaza07,golovkov09}.
Its ground state is considered to have a neutron skin structure consisting of four valence neutrons around $^4$He
with small binding energy of 3.1 MeV.
The recent experiments reported the matter and charge radius of $^8$He in addition to $^6$He \cite{tanihata92,mueller07}.
For the excited states of $^8$He, most of them can be located above the $^4$He+4$n$ threshold energy \cite{skaza07}.
This fact indicates that the observed resonances of $^8$He can decay into 
the channels of $^7$He+$n$,  $^6$He+2$n$, $^5$He+3$n$ and $^4$He+4$n$.
These multiparticle decays of $^8$He are related to the Borromean nature of $^6$He, which breaks up easily into $^4$He+$2n$,
and make it difficult to settle the excited states of $^8$He.
Similar situation is also occurred for other He isotopes, such as $^6$He and $^7$He \cite{skaza06}.

In the theoretical side, {\it ab initio} calculation of Green's function Monte Carlo \cite{pieper04}
has shown that the calculated energy levels fairly show a good correspondence with the experiments,
although the results depend on the choice of the three-nucleon forces.
This calculation is based on the bound state approximation and 
the continuum effect of the open channels is not included, 
while the excited states of $^8$He are unbound.

Several methods have been proposed to treat the continuum effects explicitly, 
such as the microscopic cluster model \cite{wurzer97,adahchour06,arai09}, the continuum shell model \cite{volya05} and the Gamow shell model \cite{betan09,michel07}.
It is, however, difficult to satisfy the multiparticle decay conditions correctly for all open channels. 
The energy spectra of many-body resonances depend on the treatment of open channels.
For $^8$He, it is necessary to describe the $^4$He+4$n$ five-body resonances in the theory.
Furthermore, it is important to reproduce the threshold energies of subsystems for particle decays.
Emphasizing these theoretical conditions, in this study, we employed the cluster orbital shell model (COSM) \cite{suzuki88,masui06}
of the $^4$He+$4n$ five-body system for He isotopes.
In COSM, the effects of all open channels are taken into account explicitly, so that we can treat the many-body decaying phenomena.
In our previous works \cite{myo077,myo09}, we have successfully obtained the $^4$He+$3n$ four-body resonances of $^7$He,
including the full couplings with $^{5,6}$He.
We have described many-body resonances as Gamow states using the complex scaling method (CSM) \cite{aoyama06,ho83,moiseyev98},
under the correct boundary conditions for all decay channels. 
In CSM, the resonant wave functions are directly obtained by 
diagonalization of the complex-scaled Hamiltonian using the $L^2$ basis functions.
The successful results of He isotopes have been obtained for energies, decay widths, spectroscopic factors, Coulomb breakups and so on.

In this study, we proceed our study of He isotopes to the $^8$He structures. 
It is interesting to see how our model describes $^8$He in addition to $^{5-7}$He and 
predicts the excited states of $^8$He.
The excited states of $^8$He can be a five-body resonance.
It is a challenge of CSM to describe the five-body nuclear resonances.
For this purpose, in this article, we concentrate on the $0^+$ states of $^8$He.
We predict the excited $0^+$ resonances and investigate their structures in comparison with the ground state.
We also calculate the monopole strength from the ground state into the unbound states of $^8$He.
This is to see the characteristics not only of the resonances, but also of non-resonant continuum states of $^8$He. 
In the breakups of $^8$He in a low energy region, two kinds of the final states of $^6$He+2$n$ and $^4$He+4$n$ are available.
Using the five-body unbound states of $^8$He obtained in COSM,
we clarify the breakup processes of $^8$He into the above final states by the monopole excitation.
Similar analysis has been performed in the three-body Coulomb breakups of halo nuclei \cite{myo01,myo03,myo0711}.


We explain the present model of He isotopes.
We use COSM of the $^4$He+$N_{\rm v} n$ systems, where $N_{\rm v}$ is a valence neutron number around $^4$He, 
namely, $N_{\rm v}=4$ for $^8$He.
The Hamiltonian is the same as that used in Refs.~\cite{myo077,myo01};
\begin{eqnarray}
	H
&=&	\sum_{i=1}^{N_{\rm v}+1}{t_i} - T_G +	\sum_{i=1}^{N_{\rm v}} V^{\alpha n}_i + \sum_{i<j}^{N_{\rm v}} V^{nn}_{ij}
        \\
&=&	 \sum_{i=1}^{N_{\rm v}} \left[ \frac{\vec{p}^2_i}{2\mu} + V^{\alpha n}_i \right] + \sum_{i<j}^{N_{\rm v}} \left[ \frac{\vec{p}_i\cdot \vec{p}_j}{4m} + V^{nn}_{ij} \right] ,
        \label{eq:Ham}
\end{eqnarray}
where $t_i$ and $T_G$ are the kinetic energies of each particle ($n$ and $^4$He) and of the center of mass of the total system, respectively.
The operator $\vec{p}_i$ is the relative momentum between $n$ and $^4$He. The reduced mass $\mu$ is $4m/5$ using a nucleon mass $m$.
The $^4$He-$n$ interaction $V^{\alpha n}$ is given by the microscopic KKNN potential \cite{aoyama06,kanada79},
in which the tensor correlation of $^4$He is renormalized into the potential 
based on the resonating group method of the $^4$He+$n$ scattering.
We use the Minnesota potential \cite{tang78} as $V^{nn}$. 
These interactions reproduce the low-energy scattering of the $^4$He-$n$ and the $n$-$n$ systems, respectively.

For the wave function, $^4$He is treated as the $(0s)^4$ configuration of a harmonic oscillator wave function, 
whose length parameter is 1.4 fm to fit the charge radius of $^4$He as 1.68 fm.
The motion of valence neutrons around $^4$He is solved variationally using the few-body technique.
We expand the relative wave functions of the $^4$He+$N_{\rm v} n$ system using the COSM basis states \cite{suzuki88,masui06}.
In COSM, the total wave function $\Psi^J$ with a spin $J$ of the $^A$He=$^4$He+$N_{\rm v} n$ system (mass number $A=4+N_{\rm v}$) is represented by the superposition of the configuration $\Psi^J_c$ as
\begin{eqnarray}
    \Psi^J(^{A}{\rm He})
&=& \sum_c C^J_c \Psi^J_c(^{A}{\rm He}),
    \label{WF0}
    \\
    \Psi^J_c(^{A}{\rm He})
&=& \prod_{i=1}^{N_{\rm v}} a^\dagger_{\alpha_i}|0\rangle, 
    \label{WF1}
\end{eqnarray}
where $^4$He corresponds to a vacuum $|0\rangle$.
The creation operator $a^\dagger_{\alpha}$ is for the single particle state of a valence neutron above $^4$He,
with the quantum number $\alpha=\{n,\ell,j\}$ in a $jj$ coupling scheme.
Here, the index $n$ represents the different radial component. 
The index $c$ represents the set of $\alpha_i$ as $c=\{\alpha_1,\cdots,\alpha_{N_{\rm v}}\}$.
We take a summation over the available configurations in Eq.~(\ref{WF0}), which give a total spin $J$.
The expansion coefficients $\{C_c^J\}$ in Eq.~(\ref{WF0}) are determined 
with respect to the total wave function $\Psi^J$ by diagonalization of the Hamiltonian matrix elements.

\begin{figure}[t]
\centering
\includegraphics[width=7.0cm,clip]{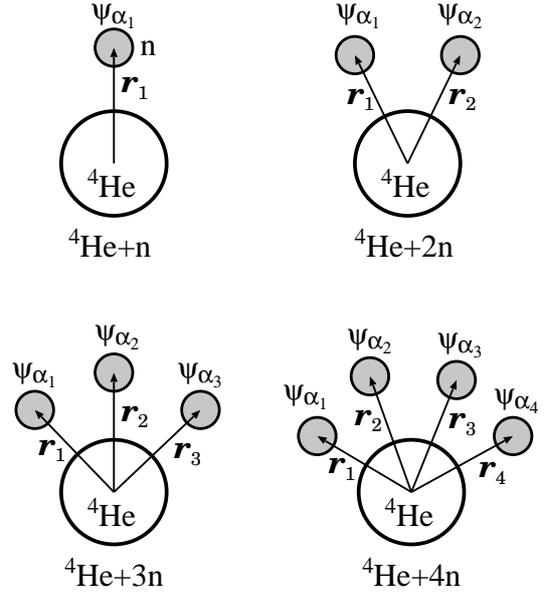}
\caption{Sets of the spatial coordinates in COSM for the $^4$He+$N_{\rm v} n$ system.}
\label{fig:COSM}
\end{figure}

The coordinate representation of the single particle state corresponding to $a^\dagger_{\alpha}$ is given as 
$\psi_{\alpha}(\vc{r})$ as function of the relative coordinate $\vc{r}$ between the center of mass of $^4$He 
and a valence neutron \cite{suzuki88,masui06,myo09}, as shown in Fig.~\ref{fig:COSM}.
We employ sufficient number of radial bases of $\psi_\alpha$
in order to describe the spatial extension of valence neutrons in the weak binding state and also in the resonances.
In this model, the radial part of $\psi_\alpha$ is expanded with Gaussian basis functions \cite{hiyama03} as
\begin{eqnarray}
    \psi_\alpha
&=& \sum_{k=1}^{N_{\ell j}} d^k_{\alpha}\ \phi_{\ell j}^k(\vc{r},b_{\ell j}^k),
    \label{WFR}
    \\
    \phi_{\ell j}^k(\vc{r},b_{\ell j}^k)
&=& {\cal N}\, r^{\ell} e^{-(r/b_{\ell j}^k)^2/2} [Y_{\ell}(\hat{\vc{r}}),\chi^\sigma_{1/2}]_{j}.
    \label{Gauss}
\end{eqnarray}
The index $k$ is for the Gaussian basis with the length parameter $b_{\ell j}^k$.
A basis number and the normalization factor of the basis are given by $N_{\ell j}$ and ${\cal N}$, respectively. 
The coefficients $\{d^k_{\alpha}\}$ in Eq.~(\ref{WFR}) are determined using the Gram-Schmidt orthonormalization,
and hence, the basis states $\psi_\alpha$ are orthogonal to each other.
The numbers of the radial bases of $\psi_\alpha$ are at most $N_{\ell j}$,
and are determined to converge the physical solutions.
The length parameters $b_{\ell j}^k$ are chosen in geometric progression \cite{aoyama06,hiyama03}.
We use at most 17 Gaussian basis functions by setting $b_{\ell j}^k$ from 0.2 fm to around 40 fm
with the geometric ratio of 1.4 as a typical one.
Due to the expansion of the radial wave function using a finite number of basis states, 
all the energy eigenvalues are discretized for bound, resonant and continuum states.
To obtain the Hamiltonian matrix elements of multi-neutron system,
we employ the $j$-scheme technique of the shell model.
The antisymmetrization between a valence neutron and $^4$He is treated on the orthogonality condition model \cite{aoyama06}, 
in which the single particle state $\psi_{\alpha}$ is imposed to be orthogonal to the $0s$ state occupied by $^4$He.

\begin{figure}[t]
\centering
\includegraphics[width=8.5cm,clip]{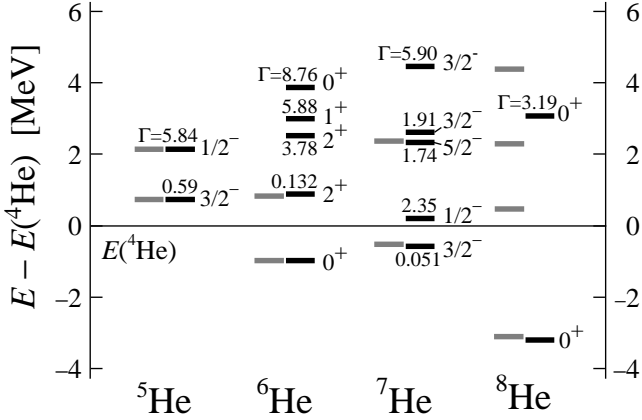}
\caption{Energy levels of He isotopes measured from the $^4$He energy. Units are in MeV.
Black and gray lines are theory and experiments, respectively. Small numbers are decay widths.
For $^8$He, the experimental data are taken from Ref. \cite{golovkov09}, and only the $0^+$ states are shown in theory.}
\label{fig:5678}
\end{figure}

In COSM, the asymptotic boundary condition of the wave functions for neutron emissions are correctly described \cite{myo09,aoyama06}.
For $^8$He, all the channels of $^8$He, $^7$He+$n$, $^6$He+$2n$, $^5$He+$3n$, $^4$He+$4n$ are automatically included in the total wave function
$\Psi^J$ in Eq.~(\ref{WF0}).
These components are coupled to each other by the interactions and the antisymmetrization,
which depend on the relative distances between $^4$He and a valence neutron and between the valence neutrons.
For the single-particle states, we take the angular momenta $\ell\le 2$ to keep the accuracy of the converged energy within 0.3 MeV 
of $^6$He in comparison with the full space calculation. 
In this model, we adjust the two-neutron separation energy of $^6$He($0^+$) to the experiment of 0.975 MeV 
by taking the 173.7 MeV of the repulsive strength of the Minnesota force instead of the original value of 200 MeV.
The adjustment of the $nn$ interaction is originated from the pairing correlation between valence neutrons with higher angular momenta $\ell>2$ \cite{aoyama06}. 
Hence, the present model reproduces the observed properties of $^{5,6}$He, as shown in Fig.~\ref{fig:5678}, 
namely, the threshold energies of the particle emissions of He isotopes.

\begin{figure}[th]
\centering
\includegraphics[width=8.5cm,clip]{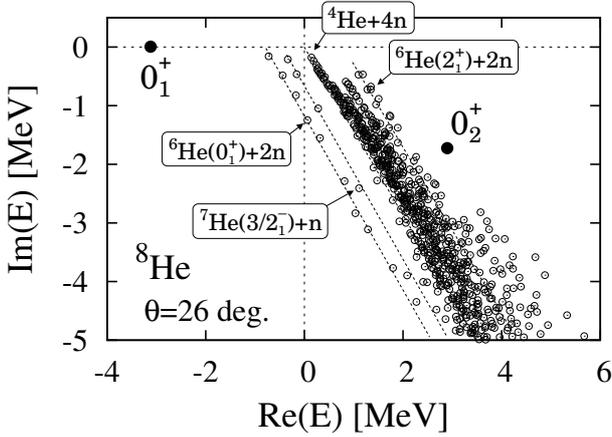}
\caption{Energy eigenvalue distribution of $^8$He($0^+$) in complex energy plane.}
\label{fig:ene8}
\end{figure}

We explain CSM, which describes resonances and nonresonant continuum states \cite{aoyama06}.
Hereafter, we refer to the nonresonant continuum states as simply the continuum states.
In CSM, we transform the relative coordinates of the $^4$He+$N_{\rm v} n$ system, as $\vc{r}_i \to \vc{r}_i\, e^{i\theta}$
for $i=1,\cdots,N_{\rm v}$, where $\theta$ is a scaling angle.
The Hamiltonian in Eq.~(\ref{eq:Ham}) is transformed into the complex-scaled Hamiltonian $H_\theta$, and the corresponding complex-scaled Schr\"odinger equation is given as
\begin{eqnarray}
	H_\theta\Psi^J_\theta
&=&     E\Psi^J_\theta,
	\label{eq:eigen}
\end{eqnarray}
The eigenstates $\Psi^J_\theta$ are obtained by solving the eigenvalue problem of $H_\theta$ in Eq.~(\ref{eq:eigen}).
In CSM, we obtain all the energy eigenvalues $E$ of bound and unbound states on a complex energy plane, governed by the ABC theorem \cite{aoyama06}.
In this theorem, it is proved that the boundary condition of Gamow resonances is transformed to the damping behavior at the asymptotic region.
This condition makes it possible to use the same method of obtaining the bound states for resonances. 
For a finite value of $\theta$, every Riemann branch cut is commonly rotated down by $2\theta$.
Hence, the continuum states such as $^7$He+$n$ and $^6$He+2$n$ channels are obtained on the branch cuts rotated with the $2\theta$ dependence \cite{myo077,myo01}. 
On the contrary, bound states and resonances are obtainable independently of $\theta$ (see Fig. \ref{fig:ene8}).
Hence, we can identify the resonance poles with complex eigenvalues: $E=E_r-i\Gamma/2$, where $E_r$ and $\Gamma$ are the resonance energies and the decay widths, respectively. 
In the wave function, the $\theta$ dependence is included in the coefficients in Eq.~(\ref{WF0}) as $\{C_c^{J,\theta}\}$. 
The value of the angle $\theta$ is determined to search for the stationary point of each resonance in a complex energy plane\cite{aoyama06,ho83,moiseyev98}.
We take $\theta$ as 20 degree in the $^8$He calculation.
In CSM, the amplitudes of the obtained resonances are finite and normalized to be unity totally, 
as $\sum_{c} \left(C_c^{J,\theta}\right)^2=1$.
Here, the Hermitian product is not applied due to the biorthogonal relation \cite{berggren68}.


In this study, we calculate the monopole strength function of $^8$He into the unbound states. 
To calculate the strength function, one needs the extended completeness relation (ECR) of $^8$He consisting of bound, resonant, and continuum states,
which are constructed using the complex-scaled eigenstates $\Psi^J_\theta$ in Eq.~(\ref{eq:eigen}).
We briefly explain ECR of $^8$He using CSM \cite{myo09,myo01,berggren68}.
When we take a large $\theta$ sufficiently, five-body unbound states of $^8$He are decomposed into several classes of the state,
which consist of the five-body ECR of $^8$He as
\begin{eqnarray}
	{\bf 1}
&=&	\sum_{~\nu} \kets{\Psi^\theta_\nu}\bras{\wtil{\Psi}^\theta_\nu}
	\nonumber
	\\
&=&	\{\mbox{bound state of $^8${He}}\}
~+~	\{\mbox{resonances of $^8${He} }\}
	\nonumber
	\\
&+&	\{\mbox{two-body continuum states of $^7${He}$^{(*)}$+$n$}\}
	\nonumber
	\\
&+&	\{\mbox{three-body continuum states of $^6${He}$^{(*)}$+$2n$}\}
	\nonumber
	\\
&+&	\{\mbox{four-body continuum states of $^5${He}$^{(*)}$+$3n$}\}
	\nonumber
	\\
&+&	\{\mbox{five-body continuum states of $^4${He}+$4n$}\} ,
	\label{eq:ECR}
\end{eqnarray}
where 
$\{ \Psi_\nu^\theta,\wtil{\Psi}_\nu^\theta \}$ forms a set of biorthogonal bases with a state $\nu$.
For simplicity, we here do not write the spin index explicitly. 
The expressions of $2n$, $3n$ and $4n$ in Eq.~(\ref{eq:ECR}) mean no-interacting states of multi-neutrons.

We explain how to calculate the strength function using ECR within CSM.
To do this, we define the complex-scaled Green's function ${\cal G}^\theta(E)$ with the energy $E$ of the system as
\begin{eqnarray}
	{\cal G}^\theta(E)
&=&	\frac{ {\bf 1} }{ E-H_\theta }
~=~	\sum_{~\nu}
	\frac{|\Psi^\theta_\nu\rangle \langle \wtil{\Psi}^\theta_\nu|}{E-E_\nu^\theta} , 
	\label{eq:green1}
\end{eqnarray}
where, the complex-scaled eigenvalue $E_\nu^\theta$ is associated with the wave function $\Psi^\theta_\nu$.
The strength function ${\cal S}_\lambda(E)$ for the operator ~$\hO_\lambda$ with rank $\lambda$
is defined in terms of Green's function without CSM as
\begin{eqnarray}
	{\cal S}_\lambda(E) 
&=&	\sum_{~\nu}
	\bras{\wtil{\Psi}_0}\hO_\lambda^\dagger\kets{\Psi_\nu}\bras{\wtil{\Psi}_\nu}\hO_\lambda\kets{\Psi_0}\
	\delta(E-E_\nu)
	\label{eq:strength_org}
	\\
&=&	-\frac1{\pi}\ {\rm Im}\left[ \langle \wtil{\Psi}_0 | \hO^\dagger_\lambda {\cal G}(E) \hO_\lambda | \Psi_0 \rangle \right] , 
	\label{eq:strength1}
\end{eqnarray}
where $\Psi_0$ is the initial state.
We operate the complex scaling on the strength function of Eq.~(\ref{eq:strength1})
and insert the complex-scaled Green's function in Eq.~(\ref{eq:green1}).
\begin{eqnarray}
	{\cal S}_\lambda(E)
&=&     \sum_{~\nu} {\cal S}_{\lambda,\nu}(E) ,
        \\
        {\cal S}_{\lambda,\nu}(E)
&=&     -\frac1{\pi}\ {\rm Im}\left[  \frac{
	\bras{\wtil{\Psi}_0^\theta}  (\hO_\lambda^\dagger)^\theta \kets{\Psi_\nu^\theta}
	\bras{\wtil{\Psi}_\nu^\theta} \hO_\lambda^\theta          \kets{\Psi_0^\theta}
        }{E-E_\nu^\theta}
        \right] . 
	\label{eq:strength3}
\end{eqnarray}
From the decomposed strength function $S_{\lambda,\nu}(E)$, we can identify the contributions of each state $\nu$ in the total strength $S_\lambda(E)$.
It is noted that the functions $S_{\lambda}(E)$ and $S_{\lambda,\nu}(E)$ are independent of $\theta$ \cite{myo09,myo01}.
This is because any matrix elements are obtained independently of $\theta$ in CSM, 
and also because the state $\nu$ of $^8$He is uniquely classified according to ECR in Eq.~(\ref{eq:ECR}).
As a result, $S_{\lambda,\nu}(E)$ is uniquely obtained.
This method has been applied to the calculations of scattering amplitudes of the breakup reactions \cite{suzuki05,kruppa07,kikuchi09}.

Here, we discuss the properties of the function $S_{\lambda,\nu}(E)$.
The total strength function $S_{\lambda}(E)$ is an observable being positive definite for every energy. 
On the other hand, the decomposed one $S_{\lambda,\nu}(E)$ is not necessarily positive definite at all energies,
because $S_{\lambda,\nu}(E)$ cannot be directly observed, similar to resonant poles.
This means that $S_{\lambda,\nu}(E)$ can sometimes have negative values.
This property of the decomposed strength has been generally discussed in Refs. \cite{myo01,myo03}.

In this study, we discretize the continuum states in terms of the basis expansion, as shown in Fig.~\ref{fig:ene8}.
The reliability of the continuum discretization in CSM has already been shown using the continuum level density \cite{suzuki05,kruppa07,kikuchi09}.



We show the obtained energy spectra of He isotopes in Fig.~\ref{fig:5678}.
One can see a good agreement between theory and experiment.
For $^6$He, the position of the $2^+$ state is good and its decay width of 0.132 MeV agrees with the experiment of 0.113(20) MeV \cite{aj89}.
For $^7$He, the ground state is located by 0.40 MeV above the $^6$He ground state,
which agrees with the recent experiments of 0.44 MeV and 0.36 MeV \cite{skaza06}.
For $^8$He, the ground state binding energy is obtained as 3.22 MeV from the $^4$He+$4n$ threshold, which agrees with 3.11 MeV of the experiment.
We predict the $0^+_2$ state with the 6.29 MeV excitation energy and the 3.19 MeV decay width.
We demonstrate the example of the $0^+$ eigenvalue distribution using CSM in Fig.~\ref{fig:ene8} with $\theta$ being 26 degree.
We successfully obtain the $0^+_2$ state of $^8$He as a five-body resonance and confirm other continuum states such as $^6$He+$2n$ and $^7$He+$n$.

The obtained matter and charge radius of $^6$He and $^8$He for their ground states are shown in Table~\ref{tab:radius} 
and reproduce the experiments.
Hence, the present model well describes the neutron halo and skin structures in He isotopes.
The proton and neutron radius are obtained as 1.82 fm and 2.60 fm for $^6$He
and 1.80 fm and 2.72 fm for $^8$He, respectively.

\nc{\lw}[1]{\smash{\lower1.5ex\hbox{#1}}}
\begin{table}[t]
\caption{Matter ($R_{\rm m}$) and charge ($R_{\rm ch}$) radius of $^6$He and $^8$He
in comparison with the experiments; a\cite{tanihata92}, b\cite{alkazov02}, c\cite{kiselev05}, d\cite{mueller07}.
Units are in fm.}
\label{tab:radius}
\centering
\small
\begin{tabular}{r|p{1.3cm} p{4.0cm}}
\hline
                     & Present  & Experiments        \\ 
\hline
\lw{$^6$He}~~~$R_{\rm m}$  &  2.37  & 2.33(4)$^{\rm a}$~~~~2.45(10)$^{\rm b}$~~~~2.37(5)$^{\rm c}$ \\
         $R_{\rm ch}$ &  2.01  & 2.068(11)$^{\rm d}$ \\
\hline
\lw{$^8$He}~~~$R_{\rm m}$  &  2.52  & 2.49(4)$^{\rm a}$~~~~2.53(8)$^{\rm b}$~~~~2.49(4)$^{\rm c}$ \\
         $R_{\rm ch}$ &  1.92  & 1.929(26)$^{\rm d}$ \\
\hline
\end{tabular}
\end{table}

\begin{table}[t]
\caption{Occupation numbers of valence neutrons in $^8$He.}
\label{tab:occupy}
\centering
\small
\begin{tabular}{p{1.0cm}|p{1.5cm} p{2.2cm}}
\hline
Orbit      &~~~~$0^+_1$  &~~~~$0^+_2$          \\ 
\hline
 $p_{1/2}$ &~~$0.14$     &~~ $1.94 -i 0.02$  \\
 $p_{3/2}$ &~~$3.71$     &~~ $2.04 -i 0.02$  \\
 $s_{1/2}$ &~~$0.02$     & $-0.02 +i 0.003$  \\
 $d_{3/2}$ &~~$0.02$     &~~ $0.04 +i 0.04$  \\
 $d_{5/2}$ &~~$0.10$     &~~ $0.01 -i 0.0004$  \\
\hline
\end{tabular}
\end{table}

\begin{figure}[b]
\centering
\includegraphics[width=8.0cm,clip]{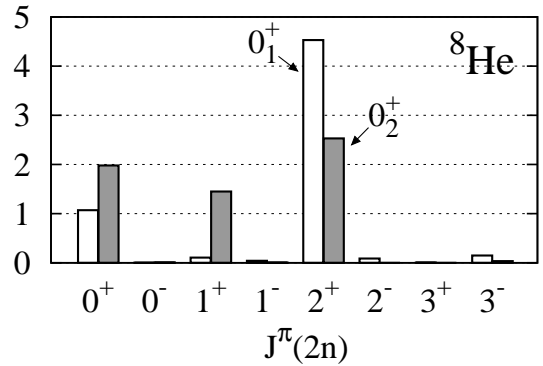}
\caption{Pair numbers of the $0^+_{1,2}$ states of $^8$He.}
\label{fig:pair}
\end{figure}

We discuss the structure of the $0^+_2$ state of $^8$He in comparison with the ground state.
We list the occupation numbers of four neutrons in each orbit for $^8$He in Tables \ref{tab:occupy}.
Since the $0^+_2$ state is a Gamow state, its quantities become complex values with a relatively small imaginary part,
while their summation conserves the valence neutron number.
In the $^8$He ground state, the $p_{3/2}$ orbit is dominant and its number is close to four.
In fact, the $(p_{3/2})^4$ configuration dominates the total wave function with a mixing of 86.0\%. 
The next dominant configurations are $(p_{3/2})^2(p_{1/2})^2$ with 6.9\%, $(p_{3/2})^2(d_{5/2})^2$ with 4.2\%,
$(p_{3/2})^2(d_{3/2})^2$ with 0.8\% and $(p_{3/2})^2(1s_{1/2})^2$ with 0.6\%.
This result means that the $jj$ coupling scheme is well established in the ground state of $^8$He.
In the $0^+_2$ state, the $p_{1/2}$ orbit is dominant with the number being around two.
In this state, the $(p_{3/2})^2(p_{1/2})^2$ configuration dominates the total wave function with a mixing of 96.9\%,
while $(p_{3/2})^4$ is given as 2.0\%.
Hence, the $0^+_2$ state of $^8$He corresponds to the $2p2h$ excited state of the ground state.

We also calculate the pair number of four valence neutrons in $^8$He, which is defined by the matrix element
of the operator $\sum_{\alpha \le \beta} A^\dagger_{J^\pi}(\alpha\beta)A_{J^\pi}(\alpha\beta)$.
Here the quantum number $\alpha$ and $\beta$ are for the single particle state and 
$A^\dagger_{J^\pi}$ ($A_{J^\pi}$) is the creation (annihilation) operator of a neutron pair with spin-parity $J^\pi$.
This pair number is useful to understand the structures of four neutrons from the viewpoint of pair coupling.
The summation of the pair number over all $J^\pi$ satisfies six from the total pair number of four neutrons.
In Fig.~\ref{fig:pair}, we show the results of the pair number up to the $3^-$ component.
In the ground state, it is found that the $2^+$ neutron pair is dominant with about 4.5 and the $0^+$ pair is almost unity.
This is consistent with the main configuration of $(p_{3/2})^4$ from the CFP decomposition ( 1 and 5 for $0^+$ and $2^+$, respectively).
The importance of the $2^+$ neutron pair is suggested in the experiment \cite{korsheninnikov03},
and is also obtained in the $^6$He+$n$+$n$ model \cite{adahchour06}.
On the other hand, the $0^+_2$ state has much $0^+$ neutron pair, about two, in addition to the large $2^+$ pair number.
This is also consistent with the $(p_{3/2})^2(p_{1/2})^2$ configuration,
which is decomposed into the pairs of $0^+$, $1^+$ and $2^+$ with the occupations of $2$, $1.5$ and $2.5$, respectively. 
The result of $0^+$ pair of neutrons in the $0^+_2$ state is interesting 
in relation with the dineutron-like cluster correlation in $^8$He suggested in AMD \cite{enyo07}.
The analysis of the spatial correlation of neutrons in the $0^+_2$ state will be performed elsewhere.


Finally, we calculate the monopole transition of $^8$He into unbound states and see the effect of the $0^+_2$ resonance,
because the monopole strength is useful to investigate the configuration properties of the states \cite{schwerdtfeger09}.
Recently, Yamada et al. \cite{yamada08} discussed the relation between the clustering excited state and its monopole strength from the ground state. 
They mentioned that the enhancement of the monopole strength can be seen in the clustering state,
because of the concentration of the strength into the relative motion of the intercluster.
In $^8$He, hence, it is interesting to investigate the monopole strength of $0^+_2$ in relation with the dineutron-like structure.
In the monopole strength, it is also important to see the effects of the continuum states in addition to that of resonance.
We take care of not only the $0^+_2$ resonance, but also all of the residual continuum states
of $^7$He+$n$, $^6$He+$2n$, $^5$He+$3n$ and $^4$He+$4n$ using ECR in Eq.~(\ref{eq:ECR}).
The angle $\theta$ in the complex scaling is taken as 20 degree to describe ECR.

\begin{figure}[t]
\centering
\includegraphics[width=8.3cm,clip]{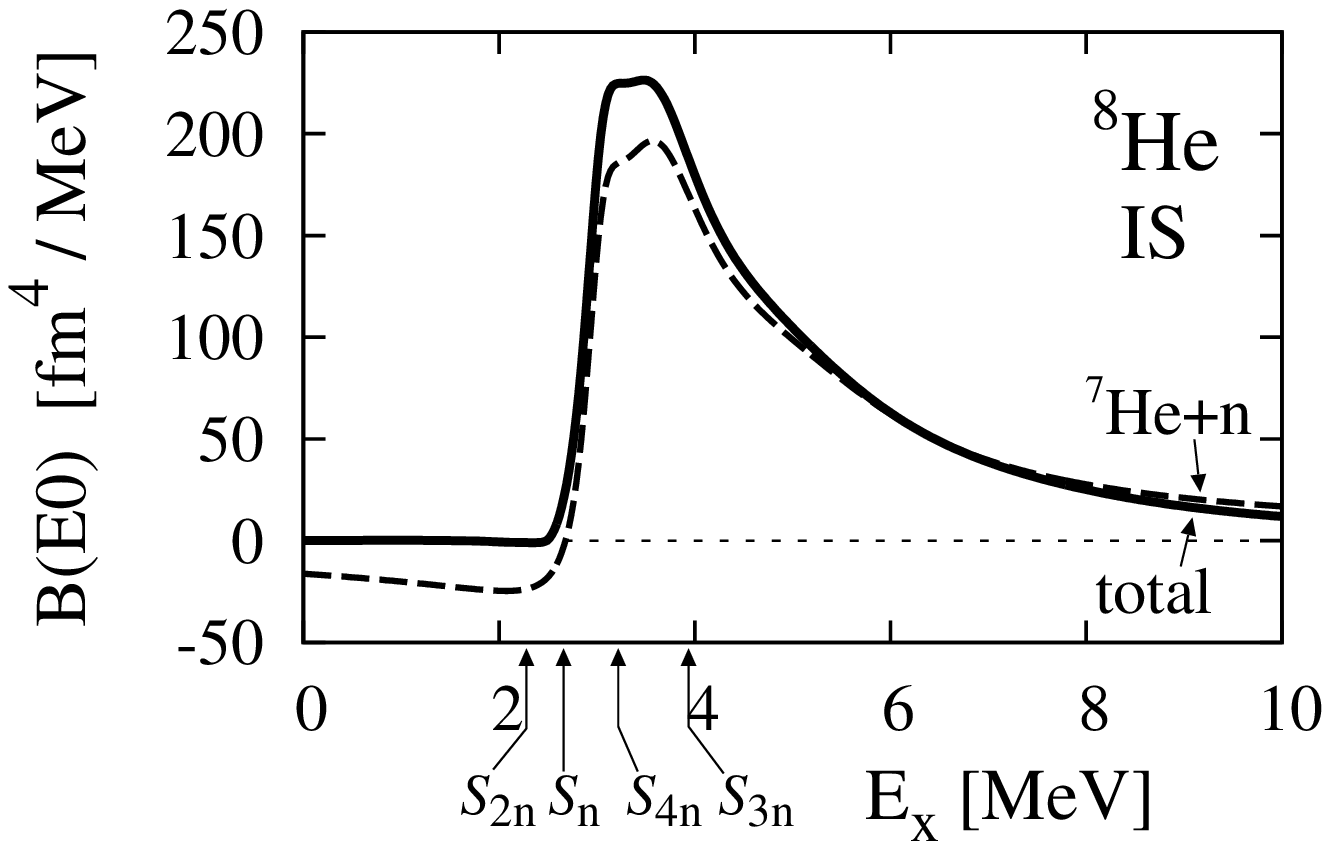}
\includegraphics[width=8.3cm,clip]{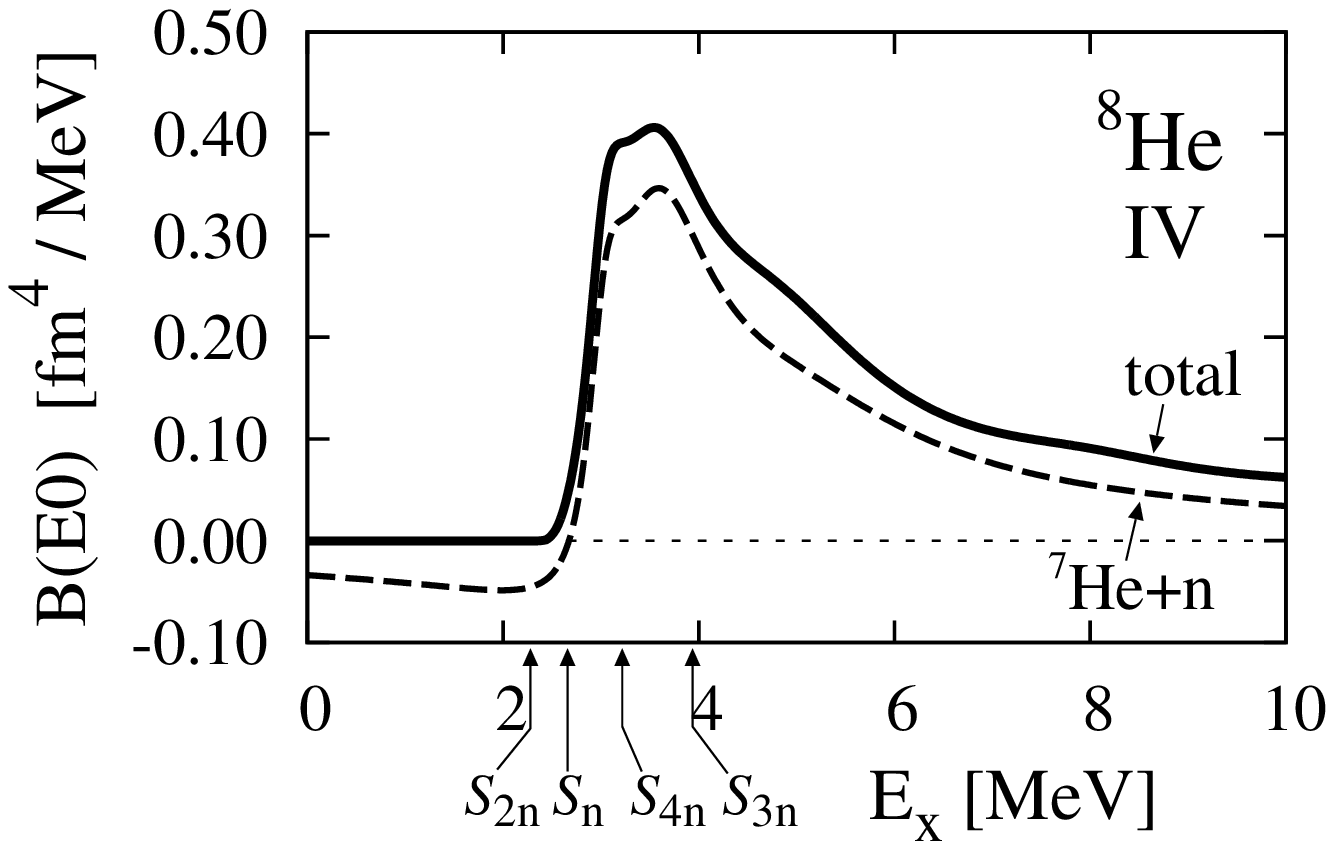}
\caption{Monopole strengths of $^8$He for isoscalar (IS) and isovector (IV) transitions as functions of the excitation energy of $^8$He.
The threshold energies of $^7$He+$n$, $^6$He+$2n$, $^5$He+$3n$ and $^4$He+$4n$ are indicated by arrows with $S_n$, $S_{2n}$, $S_{3n}$ and $S_{4n}$, respectively.}
\label{fig:monopole}
\end{figure}

In Fig.~\ref{fig:monopole}, the monopole strengths for isoscalar (IS) and isovector (IV) responses are shown.
It is found that two strengths exhibit a similar shape showing the low energy enhancement just above 3 MeV in the excitation energy.
There is no clear signature of the $0^+_2$ state around its excitation energy of 6.29 MeV in both strengths.
In fact, the transition matrix elements from the ground state into the $0^+_2$ state are
obtained as $1.78-i0.38$ fm$^4$ for IS and $-0.003+i0.018$ fm$^4$ for IV, respectively.
These values are so small in comparison with the total strengths.
This result is understood from the single particle structures of the $0^+_2$ state.
In the $0^+_2$ state, the $p_{1/2}$ orbit is largely mixed shown in Table \ref{tab:occupy},
and this orbit cannot be excited from the $p_{3/2}$ orbit in the ground state by the monopole operator.
As a result, the monopole strength into $0^+_2$ becomes negligible. 
Instead, the continuum strength gives a main contribution, which makes it difficult to observe the $0^+_2$ state via the monopole transition.
Our results do not support the enhancement of the strength into the $0^+_2$ state, however,
the possibility of the dineutron-like structure in $^8$He should be investigated carefully, in addition to the monopole strength.
It is also necessary to search for the observables which are responsible for the $0^+_2$ state.
Since the present $^8$He wave functions contain not only resonances but also continuum states,
the application to the sophisticated reaction analysis including continuum coupling would be promising.
One of the candidates of the reactions is considered to be the two-neutron transfer into $^6$He to produce the excited states of $^8$He.
Experimentally, the $^6$He($t$,$p$)$^8$He reaction was reported \cite{golovkov09}
and the observed cross section shows some peaks without spin assignment, around the resonance energy of $0^+_2$ obtained in this study.

We decompose the monopole strengths in Fig.~\ref{fig:monopole} into several continuum components using ECR and see the individual contribution.
From the analysis, it is found that the IS and IV strengths both dominantly come from the $^7$He($3/2^-_1$)+$n$ components.
This selectivity of the continuum states is related with the properties of the monopole operator, which are one-body concerning with $\vc{r}_i$ 
in Fig. \ref{fig:COSM}.
By the monopole operator, one of the relative motions of $^8$He can be strongly excited. 
As a result, the intercluster motion between the $^7$He cluster and a valence neutron is strongly coupled
with the ground state by the monopole excitation.
The obtained result also indicates the sequential breakup process of $^8$He(G.S.) $\to$ $^7$He($3/2^-_1$)+$n$ $\to$ $^6$He(G.S.)+$n$+$n$ in the monopole excitation.
Experimentally, the large contribution of the sequential process via $^7$He+$n$ was also reported in the Coulomb breakup of $^8$He \cite{iwata00},
which is dominated by the $E1$ transition.

It is interesting to see the components of the final states of $^6$He+$n$+$n$ and $^4$He+$n$+$n$+$n$+$n$ in the monopole strengths. 
In Table \ref{tab:ratio}, the ratios of the integrated strengths into two final states with respect to the total ones are shown.
Dominance of the $^6$He+$n$+$n$ state is commonly found in the IS and IV cases and more significant in the IS case.
This is because one of the relative coordinates of $\{\vc{r}_i\}$ in $^8$He can be excited independently by the IS response, and then
the $^7$He component largely remains and decays into $^6$He+$n$.
In the IV case, protons are included only in $^4$He, so that the IV response excites the relative motion between the $^4$He core and the center of 
mass of $4n$ as a recoil effect.
Owing to the excitation of the relative motion, the components of $^6$He and $^7$He in $^8$He are relatively dissolved than the IS case 
and the transition into $^4$He+$n$+$n$+$n$+$n$ increases in the IV response.

\begin{table}[t]
\caption{Ratios of the integrated monopole strengths into $^6$He+$n$+$n$ and $^4$He+$n$+$n$+$n$+$n$ final states.}
\label{tab:ratio}
\centering
\small
\begin{tabular}{r p{2.0cm} p{2.0cm}}
\hline
    & $^6$He+$n$+$n$ & $^4$He+$n$+$n$+$n$+$n$  \\ 
\hline
IS  &  0.939          &  0.061    \\
IV  &  0.690          &  0.310    \\
\hline
\end{tabular}
\end{table}

In summary, we have investigated the structures of the $0^+$ states of $^8$He in a five-body cluster model.
The boundary condition for many-body resonances is accurately treated using CSM.
We successfully obtain the five-body $0^+_2$ resonance of $^8$He in CSM.
This state dominantly has a $(p_{3/2})^2(p_{1/2})^2$ configuration and mainly consists of two of the $0^+$ neutron pairs.
We further investigate the monopole strengths into five-body unbound states,
which are described by using ECR within CSM.
It is found that the $0^+_2$ contribution is negligible in the strength,
so that it is difficult to observe the $0^+_2$ state from the monopole strength.
It is dominant that the sequential breakup process of $^8$He via the $^7$He+$n$ states into the $^6$He+$2n$ three-body final states,
instead of the $^4$He+$4n$ five-body states.

\section*{Acknowledgments}
We would like to thank Prof. Kiyomi Ikeda for fruitful discussions. 
This work was supported by a Grant-in-Aid for Young Scientists from the Japan Society for the Promotion of Science (No. 21740194).
Numerical calculations were performed on a supercomputer (NEC SX8R) at RCNP, Osaka University.

\end{document}